\documentclass[%
 reprint,longbibliography,preprintnumbers,
nofootinbib,
 amsmath,amssymb,
 aps
]{revtex4-1}
\pdfoutput=1
\usepackage{graphicx}
\usepackage[utf8]{inputenc}
\usepackage{flushend}
\usepackage{dcolumn}
\usepackage{bm}
\usepackage{balance}

\usepackage[normalem]{ulem}

\usepackage[colorlinks = true,
            linkcolor = blue,
            urlcolor  = blue,
            citecolor =green,
            anchorcolor = blue]{hyperref}
\usepackage{verbatim}
\usepackage{color,ulem}
\usepackage[english]{babel}

\usepackage[utf8]{inputenc}
\input Starburst.fd
\newcommand*\initfamily{\usefont{U}{Starburst}{xl}{n}}\initfamily

\newcommand{\beq}{\begin{eqnarray}}
\newcommand{\eeq}{\end{eqnarray}}
\usepackage{amsmath}
\usepackage{tikz}
\usetikzlibrary{decorations.pathmorphing}
\usetikzlibrary{shapes.misc}
\tikzset{cross/.style={cross out, draw=black, minimum size=8*(#1-\pgflinewidth), inner sep=0pt, outer sep=0pt},
cross/.default={1pt}}
\usetikzlibrary{patterns,math}
\begin{document}

\title{\Large Reply to R. Blumenfeld on ``Explicit analytical solution for random close packing in $d=2$ and $d=3$''}

\author{\textbf{Alessio Zaccone}$^{1}$}%
 \email{alessio.zaccone@unimi.it}
 
 \vspace{1cm}
 
\affiliation{$^{1}$Department of Physics ``A. Pontremoli'', University of Milan, via Celoria 16,
20133 Milan, Italy.}


\maketitle
In a quick response to our recent work \cite{Zaccone2022} on an analytical derivation of the random close packing (RCP) density in $d=2$ and $d=3$ based on statistical arguments due to liquid-theory combined with marginal stability \cite{Scossa}, Raphael Blumenfeld (RB) \cite{Blumenfeld_Comment} argues that our analytical solution is not consistent with RB's recent geometric approach (cfr. \cite{Blumenfeld_Comment} and references therein). 
RB argues that the value of RCP in $d=2$ obtained in \cite{Zaccone2022}, i.e. $\phi_{RCP}=0.886$, once it is plugged into his heuristic approach, generates values of polygon order that are indicative of significant polycrystallinity.

Plugging our value $\phi_{RCP}=0.88644$ from \cite{Zaccone2022} into his numerical formula, derived with several approximations that can hardly be verified, RB then deduces the corresponding mean cell order $\Bar{k}$, and argues that this value is linked to a mean coordination number $z$ that is larger than $z=4$ in $d=2$. We were not able to reproduce RB's elaborate calculations, and we cannot verify whether RB's method is valid. 

On a more epistemological level, even if RB's heuristic approach were accurate, this result would not bring new information and appears in contradiction with earlier results.
First of all, the fact that RB's peculiar protocol (many others exist) yields some structural order at $\phi_{RCP}=0.88644$ does not necessarily imply that, according to other protocols or methods of analysis, there could be packings which are fully amorphous at $\phi_{RCP}=0.88644$. This is because of several reasons, first of all because RB's approach contains several assumptions and approximations.
Furthermore, an earlier algorithm by Makse and co-workers \cite{Makse_2010} yields a value $\phi_{RCP}=0.89$ that is very close to the one we obtained in \cite{Zaccone2022}, and in fact even larger, which already disproves RB's analysis of our result.

RB's arguing in terms of loose random packing (RLP) for frictionless packings is misleading, see e.g. \cite{Torquato_review}, as in the literature RLP has been used predominantly in the context of frictional packings. Also, it is incorrect, since even classic experiments agree that $z=6$ for RCP in $d=3$ \cite{Bernal}.

Contrary to RB's line of thought, cfr. \cite{Blumenfeld_Comment} and references therein, there is no ``exact'' value for RCP in $d=2$.
A more meaningful question would be to ask what is the maximum packing fraction after which crystallinity sets in. The proper way to answer this question would be to define reliable order metrics, which is still an open question. The usual metrics based on bond-orientational order parameters, $q_{6}$ or $F_{6}$, proposed in \cite{Truskett} are far from being optimal in this sense, as they cannot capture several key features of disordered or partly crystalline systems. For example, bond-orientational order parameters are unable to correlate with the boson peak in the vibrational density of states \cite{Milkus}. 

A sensitive analysis of crystallinity and its onset was performed in \cite{Yuliang} for random sphere packings in $d=3$, and it was found that the onset of crystallinity occurs rather abruptly with a discontinuous jump reminiscent of a first-order phase transition. Interestingly, in Figs. 8 and 10 of \cite{Yuliang}, the onset of crystallinity in $d=3$ occurs at a packing fraction $\phi\approx 0.658$, which strikingly coincides with the value derived in our work \cite{Zaccone2022}.

Even more persuasively, the estimated entropy of the packings in Fig. 11(a) of Ref. \cite{Yuliang} shows a kink right at $\phi\approx 0.658$, i.e. again the value predicted in \cite{Zaccone2022}.
Although unfortunately no such analysis is available for $d=2$, where the situation is significantly complicated by the possible occurrence of  Berezinskii-Kosterlitz-Thouless (BKT) type transitions with a smoother onset of crystallinity, the above evidence in $d=3$ supports the analytical solution of \cite{Zaccone2022}.

RB argues that liquid theories are not accurate near RCP and that this is the source of ``error'' in \cite{Zaccone2022}, without bringing any mathematical evidence of what this ``error'' could be. In \cite{Zaccone2022} there are no mathematical errors, as anyone can verify, and the fact that liquid theories lack accuracy was even pointed out, very clearly, by us in our paper \cite{Zaccone2022}, in the introduction. What is interesting, and worth investigating, is that liquid theory seems to contain the RCP density even when one would expect it not to, as already discussed earlier by other authors \cite{Kamien}.

\bibliographystyle{apsrev4-1}

\bibliography{refs}

\end{document}